\documentclass[runningheads]{llncs}
\usepackage{graphicx}
\usepackage{hyperref}
\usepackage{url}
\usepackage{pdfpages}
\usepackage{fnpct}
\usepackage{booktabs}
\usepackage{multirow}
\usepackage{array}
\usepackage{tipa}
\usepackage{todonotes}
\usepackage{bibnames}
\usepackage[T4,T1]{fontenc}
\usepackage[utf8]{inputenc}
\usepackage[normalem]{ulem}

\newcommand{\phz}{\phantom{0}}

\newcommand{\AS}[1]{{\fontencoding{T4}\selectfont#1}}
\begin{document}

\title{Multilingual training set selection for ASR in under-resourced Malian languages}

\author{Ewald van der Westhuizen\orcidID{0000-0002-7430-503X} \and
Trideba Padhi\orcidID{0000-0002-4544-5770} \and
Thomas Niesler\orcidID{0000-0002-7341-1017}}

\authorrunning{E. van der Westhuizen et al.}

\institute{Department of Electrical and Electronic Engineering, Stellenbosch University, Stellenbosch, South Africa\\
\email{\{ewaldvdw,tpadhi,trn\}@sun.ac.za}}

\maketitle

\begin{abstract}
We present first speech recognition systems for the two severely under-resourced Malian languages Bambara and Maasina Fulfulde.
These systems will be used by the United Nations as part of a monitoring system to inform and support humanitarian programmes in rural Africa.
We have compiled datasets in Bambara and Maasina Fulfulde, but since these are very small, we take advantage of six similarly under-resourced datasets in other languages for multilingual training.
We focus specifically on the best composition of the multilingual pool of speech data for multilingual training.
We find that, although maximising the training pool by including all six additional languages provides improved speech recognition in both target languages, substantially better performance can be achieved by a more judicious choice.
Our experiments show that the addition of just one language provides best performance.
For Bambara, this additional language is Maasina Fulfulde, and its introduction leads to a relative word error rate reduction of 6.7\%, as opposed to a 2.4\% relative reduction achieved when pooling all six additional languages.
For the case of Maasina Fulfulde, best performance was achieved when adding only Luganda, leading to a relative word error rate improvement of 9.4\% as opposed to a 3.9\% relative improvement when pooling all six languages.
We conclude that careful selection of the out-of-language data is worthwhile for multilingual training even in highly under-resourced settings, and that the general assumption that more data is better does not always hold.

\end{abstract}

\section{Introduction}
Radio phone-in talk shows have been found to be a popular platform for voicing concerns and views regarding social issues in societies who do not have ready access to the internet.
The United Nations (UN) frequently heads humanitarian programmes in regions on the African continent where this is the case.\footnote{\scriptsize
	\texttt{https://www.unglobalpulse.org/project/\newline making-ugandan-community-radio-machine-readable-\newline using-speech-recognition-technology/}
	\texttt{https://www.unglobalpulse.org/document/using-machine-\newline learning-to-analyse-radio-content-in-uganda/}
}
In order to inform and support these efforts, a radio browsing system has been developed with the UN.
These ASR-based systems have successfully been piloted in Uganda in three local languages: Luganda, Acholi and English~\cite{menon2017radio,Saeb2017}.
Recently, a focus of UN humanitarian relief efforts has arisen in the West African country of Mali.
In this paper we describe our first ASR systems for the severely under-resourced Malian languages Bambara and Maasina Fulfulde.
This includes the compilation of annotated radio speech corpora in both languages.
Since these new resources are very limited and are time-consuming, expensive and difficult to produce, we incorporate resources from other languages that were at our disposal in various combinations for multilingual acoustic model training~\cite{katzner2002languages,schultz2006multilingual,schultz2001language}.

Multilingual training of deep neural network (DNN) acoustic models includes multiple languages in the training pool and exploits common phonetic structures among the languages.
Employing the hidden layers of a DNN as a multilingually-trained universal feature extractor has been shown to achieve good ASR performance~\cite{heigold2013dnn,huang2013dnn,vesely2012language}.
Model adaptation and transfer learning approaches from other well resourced languages are also popular approaches to acoustic model training in low-resource settings.
Here, the initial layers of a DNN trained on a well resourced language are retained, while the output layer is retrained using the smaller dataset in the target language~\cite{grezl2014adaptation}.
Finally, multilingually-trained bottleneck features have also been shown to benefit ASR performance for severely under-resourced languages~\cite{Padhi2020,vesely2012language}.

When embarking on multilingual training, an important decision is which languages to include in the training pool.
The addition of a language can sometimes lead to an increased mismatch among the datasets that leads to a deterioration in ASR performance.
Hence, the specific choice of languages in the training pool becomes a key factor in achieving the best performing ASR system.

In the next section we give an overview of the target languages.
In Section~\ref{sec:corpora}, the speech and text resources are introduced.
The description of the pronunciation, language and acoustic modelling follows in Sections~\ref{sec:pronmodelling}, ~\ref{sec:langmodelling} and \ref{sec:acousticmodelling}.
The experimental results and the discussion thereof are presented in Section~\ref{sec:results} and we conclude in Section~\ref{sec:conclusions}.

\section{Languages}

\begin{figure}
\centering
\includegraphics[width=0.98\linewidth]{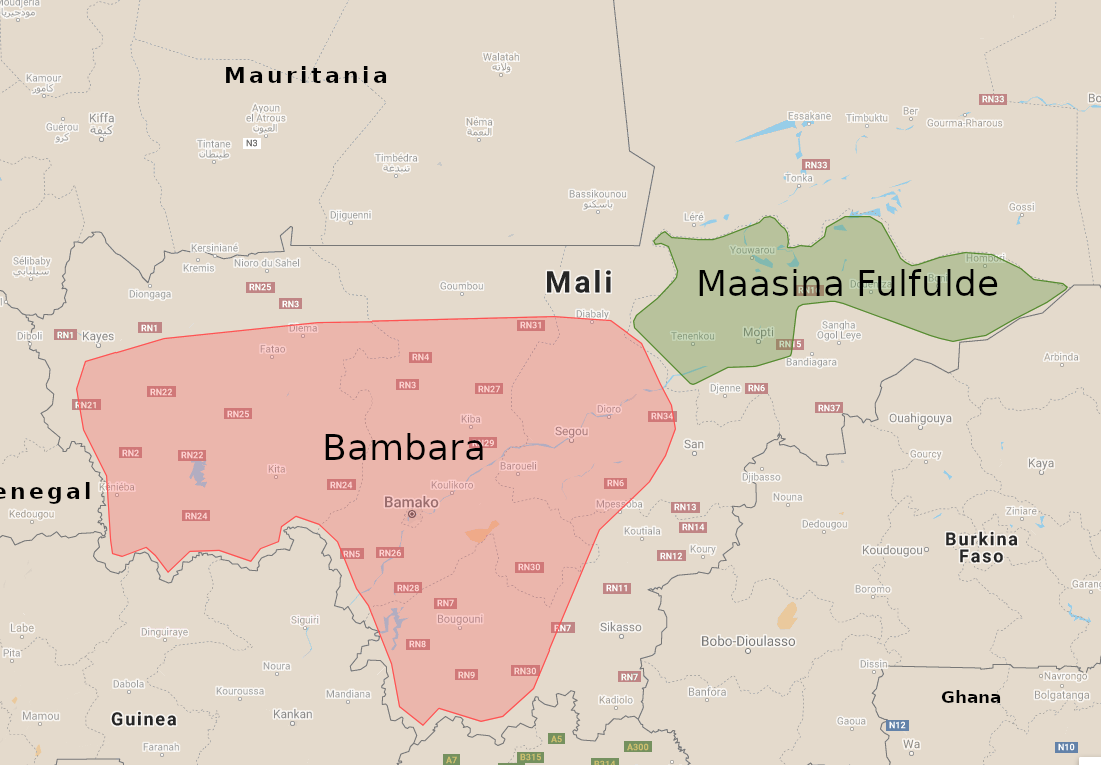}
\caption{
A map showing the regions of Mali where Bambara and Maasina Fulfulde are predominantly spoken. Adapted from Google Maps.
}
\label{fig:bam_ful_map}
\end{figure}

Mali has a multilingual and multiethnic population with 14 officially recognised national languages.
We will focus on two of these: Bambara and Maasina Fulfulde.
Figure~\ref{fig:bam_ful_map} presents a map indicating where these languages are predominantly spoken.

\subsection{The Bambara language}

Bambara is part of the Manding language group which belongs to the greater Mande and Niger-Congo language families~\cite{Vydrine2015}.
The Manding group is considered to comprise a continuum of languages and dialects that are mutually intelligible.
Bambara is used as a \textit{lingua franca} by approximately 14 million Malians.

Bambara is an agglutinative language with seven vowels, 21 consonants and a syllabic nasal~\cite{Donaldson2017b}.
Three distinct alphabets are used to write Bambara: Arabic, Latin and N'Ko~\cite{Donaldson2017b,Donaldson2017a,Vydrine2015}.
The Latin-based alphabet is officially recognised in Mali.
Bambara is a tonal language and diacritics can be used to indicate the low and high vowel tones.

\subsection{The Maasina Fulfulde language}

Maasina Fulfulde is part of the Fulfulde language group which in turn forms part of the Niger-Congo language family.
The Fulfulde language group is a continuous chain of dialects that are mutually intelligible and are spoken predominantly in West Africa and a few parts of the Central African region.
Maasina Fulfulde is spoken in the central part of Mali in the region of Macina.
It is one of Mali's official languages and has an estimated one million speakers~\cite{fagerberg1984}.

Maasina Fulfulde has 23 consonants, four prenasalised consonants and five vowels~\cite{Osborn1993}.
Unlike most of its neighbouring African languages, Fulfulde is not tonal.
Instead, it is an intonational language, where the main outlines of sentential pitch contours are determined by the sentence type, rather than by the tonal characteristics of individual words and complexes \cite{Arnott-1970,arnott1974some}.
As for Bambara, three alternative alphabets are in use for Maasina Fulfulde: Arabic, Latin and more recently Adlam~\cite{barry2018proposal}.
However, in contrast to Bambara, there is currently no universally accepted standard script.

\subsection{Scripts}
\label{sec:scripts}

As mentioned in the previous section, both Bambara and Maasina Fulfulde are written in more than one type of script.
In many African languages, the choice of script and the standardisation of orthography remains unresolved.
In the case of Manding languages, it has been pointed out that these choices are influenced by sociopolitical motivations~\cite{Donaldson2017a}.
However, in ASR, consistent orthography is a necessity, as it ensures the regular mapping of the orthographic transcriptions to the phonemic representation used for acoustic modelling.
For both Bambara and Fulfulde, the Latin-based orthography exhibits a highly regular letter-to-sound correspondence~\cite{Donaldson2017b}.
This has the advantage that pronunciation modelling becomes straightforward because graphemic units can be used, sidestepping the need to develop a pronunciation dictionary which requires highly-skilled linguistic expertise that is difficult to obtain.
Hence, in our corpus development we have chosen to use the Latin script for both languages.

\section{Speech and text corpora}
\label{sec:corpora}

No speech or text resources were available for Bambara or Maasina Fulfulde.
In this section we describe the corpora that were compiled as part of this work in order to develop acoustic and language models for use in the radio browsing systems.

\subsection{Monolingual speech corpora}
\label{sec:mono_corpora}

Our Bambara and Maasina Fulfulde speech corpora are under active development.
Transmissions from public broadcast radio are recorded and stored as 5-minute segments.
The speech in these recordings is manually segmented and transcribed by first-language speakers using Praat~\cite{Boersma2021}.
We chose to use the Latin script for both languages, as shown in Table~\ref{tab:bam_ful_alphabets}.
Although Bambara and Fulfulde are tonal and intonational languages, respectively, tone markers were not included in order to simplify and expedite the transcription process.
At the time of writing, 180 Bambara and 190 Fulfulde recordings have been transcribed, yielding approximately 10 hours of speech in each language.
Table~\ref{tab:mono_corpus_stats} summarises the extent of the training, development and test sets into which the two corpora have been partitioned.

\begin{table}[t]
	\renewcommand{\arraystretch}{1.1}
	\footnotesize
	\centering
	\caption{
		Alphabets used to transcribe Bambara and Fulfulde.
	}
	\label{tab:bam_ful_alphabets}
	\begin{tabular*}{0.97\linewidth}{@{\extracolsep{\fill}} ll @{}}
		\toprule
		\multicolumn{2}{c}{\textbf{Bambara}} \\
		\midrule
		Consonants & b, c, d, f, g, h, j, k, l, m, n, \textltailn, \AS{\ng{}}, p, r, s, t, u, w, y, z \\
		Vowels & a, e, \textepsilon, i, o, \textopeno, u \\
		\midrule
		\multicolumn{2}{c}{\textbf{Maasina Fulfulde}} \\
		\midrule
		\multirow{2}{*}{Consonants} & ', b, \texthtb, c, d, \texthtd, f, g, h, j, k, l, m, mb, n, nd, ng, nj, \\
		& \AS{\ng{}}, \textltailn, p, r, s, t, w, y, \AS{\m{y}} \\
		Vowels & a, e, i, o, u \\
		\bottomrule
	\end{tabular*}
\end{table}

\begin{table}[t]
	\footnotesize
	\caption{
		Training, development and test sets for the Bambara and Maasina Fulfulde radio speech corpora.
		(m: minutes; h: hours.)
	}
	\centering
	\begin{tabular*}{0.97\linewidth}{@{\extracolsep{\fill}}lrr @{}}
		\toprule
		\textbf{Bambara} & \textbf{Utterances} & \textbf{Duration} \\
		\midrule
		Training    & 14\,037 & 8.7 h \\
		Development &  1\,177 & 49.5 m \\
		Test        &     933 & 43.5 m \\[1mm]
		Total       & 16\,147 & 10.3 h \\
		\bottomrule
		\toprule
		\textbf{Maasina Fulfulde} & \textbf{Utterances} & \textbf{Duration} \\
		\midrule
		Training    & 14\,433 & 10.0 h \\
		Development &     322 & 7.7 m \\
		Test        &  1\,479 & 31.4 m \\[1mm]
		Total       & 16\,234 & 10.6 h \\
		\bottomrule
	\end{tabular*}
	\label{tab:mono_corpus_stats}
\end{table}

\subsection{Multilingual speech corpora}

Besides the transcribed Bambara and Maasina Fulfulde corpora desribed above, we had at our disposal speech corpora in a number of other languages.
These include radio speech datasets collected for the previously-developed radio browsing systems in Luganda, Acholi and Ugandan English~\cite{menon2017radio}.
The data collection procedure for these three languages was similar to that employed for Bambara and Maasina Fulfulde, thus providing a degree of uniformity.
In addition, we included Arabic\footnote{OpenSLR Tunisian Modern Standard Arabic corpus, accessed 2021-02-21 at \url{http://www.openslr.org/46/}}, since it is a national language of Mali.
Finally, we have collected a small corpus of radio speech in Malian French, which is also a national language of Mali.
However, inclusion of this data led to consistent deterioration in preliminary experiments and hence it will not be reported on.
The statistics for the four corpora used for multilingual training are summarised in Table~\ref{tab:multi_corpus_stats}.

\begin{table}[h]
	\footnotesize
	\centering
	\caption{
		Additional speech corpora used for multilingual acoustic modelling.
		Duration in hours.
	}
	\label{tab:multi_corpus_stats}
	\begin{tabular*}{0.97\linewidth}{@{\extracolsep{\fill}}l @{\hspace{3pt}} r @{\hspace{3pt}} r @{}}
		\midrule
		\textbf{Corpus} & \textbf{Utterances} & \textbf{Duration} \\
		\midrule
		Luganda                         &  9\,001 &  9.8 \\
		Acholi                          &  4\,860 &  9.2 \\
		Ugandan English                 &  4\,402 &  5.7 \\
		Tunisian Modern Standard Arabic & 11\,688 & 11.2 \\
		\midrule
		Total                           & 29\,951 & 35.9 \\
		\bottomrule
	\end{tabular*}
\end{table}

\subsection{Text corpora}
\label{sec:text_corpora}

\begin{table}[t]
	\footnotesize
	\centering
	\caption{
		Text corpora collected for Bambara and Maasina Fulfulde.
	}
	\label{tab:text_stats}
	\begin{tabular*}{0.97\linewidth}{@{\extracolsep{\fill}}l @{\hspace{3pt}} r @{\hspace{3pt}} r @{}}
		\midrule
		\textbf{Corpus}     & \textbf{Types} & \textbf{Tokens} \\
		\midrule
		Bambara             &  19\,992       &  553\,640 \\
		Maasina Fulfulde    &  25\,637       &  665\,775 \\
		\bottomrule
	\end{tabular*}
\end{table}

Digital text resources in Bambara and Maasina Fulfulde have been compiled from the internet, and this process of identification and collection of additional text is ongoing.
To the best of our knowledge, no freely or commercially available resources exist in these severely under-resourced languages.
The corpora in Table~\ref{tab:text_stats} represent the full extent of text available for language modelling in addition to the training transcriptions described in Table~\ref{tab:mono_corpus_stats}.
Sources include Wikipedia, newspaper articles, blog posts and comments and religious texts.

\section{Pronunciation modelling}
\label{sec:pronmodelling}

As discussed in Section~\ref{sec:scripts}, Bambara and Fulfulde have highly regular letter-to-sound correspondence.
We have therefore constructed the pronunciation dictionary using graphemes.
This is a simple, fast and automated process that does not rely on skilled linguistic expertise.
For example, the dictionary entry for the Bambara word, \emph{ang\textepsilon r\textepsilon},\ is:\newline
\texttt{\hspace*{10pt}\footnotesize ang\textepsilon r\textepsilon\  a\textunderscore bam n\textunderscore bam g\textunderscore bam \textepsilon\textunderscore bam r\textunderscore bam \textepsilon\textunderscore bam}\newline
A language tag is appended to each grapheme-based phoneme label to disambiguate the labels from similar symbols that may exist in other languages in the multilingual training pool.
The same process is applied to create the grapheme-based pronunciation dictionary for Fulfulde.
The pronunciation dictionaries for the other languages in the multilingual training pool (Table~\ref{tab:multi_corpus_stats}) have been manually compiled by language experts and use true phoneme labels.
A multilingual pronunciation dictionary is constructed by concatenating all the dictionaries of the languages in the mulitlingual training pool.

\section{Language modelling}
\label{sec:langmodelling}

The vocabularies of the language models were closed to include the word types in the training, development and test sets.
The vocabularies of the Bambara and Fulfulde language models contain 15\,526 and 12\,368 word types respectively.

The SRILM toolkit~\cite{Stolcke2002} was used to train trigram language models.
Preliminary experiments indicated that higher order models did not provide any further benefit.
The same training procedure was used to obtain the language models in each of our two target languages.
First, a baseline language model is trained using the training transcriptions (Table~\ref{tab:mono_corpus_stats}).
Thereafter, language models trained on the additional text resources (Table~\ref{tab:text_stats}) are interpolated with the baseline to obtain the final language model that is used in the ASR experiments.
The interpolation weights were chosen to optimise the development set perplexity.
The test set perplexities for the baseline and interpolated language models are given in Table~\ref{tab:lmppls}.
The fairly small perplexity improvements (<5\%) indicate a mismatch between the language style in the speech transcripts and the additional text resources.

\begin{table}[t]
	\renewcommand{\arraystretch}{1.2}
	\footnotesize
	\centering
	\caption{Perplexity (PP) of the baseline (transcriptions only) and interpolated (transcriptions and additional text) language models.}
	\label{tab:lmppls}
	\begin{tabular*}{0.99\linewidth}{@{\extracolsep{\fill}}p{130pt}rr @{}}
		\toprule
		\textbf{Bambara} & \textbf{Word tokens} & \textbf{PP\phz} \\
		\midrule
		Baseline language model       & 137\,250 & 202.46 \\
		Interpolated language model   & 690\,890 & 193.33 \\
		\bottomrule
		\toprule
		\textbf{Maasina Fulfulde}     & \textbf{Word tokens} & \textbf{PP\phz} \\
		\midrule
		Baseline language model       & 135\,054 & 522.59 \\
		Interpolated language model   & 800\,809 & 504.56 \\
		\bottomrule
	\end{tabular*}
\end{table}

\section{Acoustic modelling}
\label{sec:acousticmodelling}

\begin{figure}
	\centering
	\includegraphics[width=0.55\linewidth]{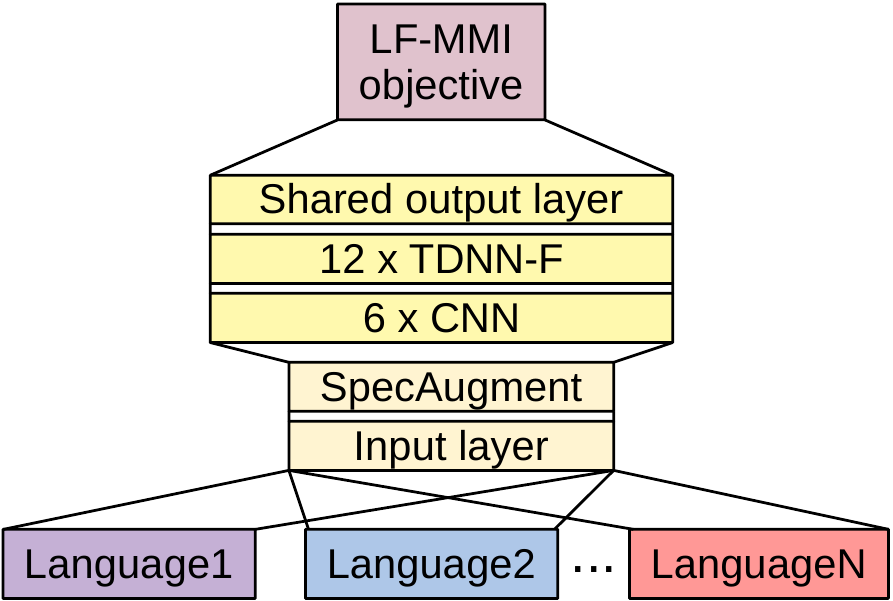}
	\caption{Multilingual training DNN architecture.}
	\label{fig:multilingual_training}
\end{figure}

The acoustic models were trained and evaluated using Kaldi~\cite{povey2011kaldi}.
We followed the standard practice of training a hidden Markov model/Gaussian mixture model (HMM/GMM) system to obtain alignments for DNN training.
The DNN architecture used for the multilingual acoustic modelling follows the Librispeech recipe and consists of six convolutional neural network (CNN) layers followed by twelve factorised time-delay deep neural network (TDNN-F) layers, as shown in Figure~\ref{fig:multilingual_training}.
The hidden layers and the output layer are shared between languages and the LF-MMI criterion is used as training objective~\cite{Povey2016}.
A \emph{SpecAugment} layer precedes the CNN layers as it has proved beneficial in all experiments~\cite{Park2019}.
Each ASR system is trained from scratch for six epochs from a randomly initialised network.
Three-fold speed perturbation is used during training~\cite{Ko2015}.

In our experiments, we investigate how the composition of the multilingual training pool affects ASR performance.
With no prior knowledge about the relatedness of the corpora and languages, a reasonable approach might be to pool all the data that we have at our disposal.
This training set composition of six pooled languages is indicated by ALL.
We also considered several possible subsets of the six languages for multilingual training, with the constraint that the target language was always included.
In each case, a multilingual model is trained on the selected pool of languages, after which it is evaluated on the target language.
Adaptation subsequent to multilingual training was attempted in several ways, but in all cases led to deteriorated performance.
Similar observations have for example been made for Ethiopian languages~\cite{Tachbelie2020}.
For computational reasons, we did not explore all the language combinations exhaustively.
We compare the speech recognition performance of these multilingually-trained systems with each other and with the baseline monolingual systems (denoted Bmono and Fmono) trained on only the target language corpora.

\section{Results and Discussion}
\label{sec:results}

\begin{figure}[!t]
\centering
\begin{tabular}{m{50pt}m{380pt}}
\raggedleft(a) & \includegraphics[width=0.6\linewidth]{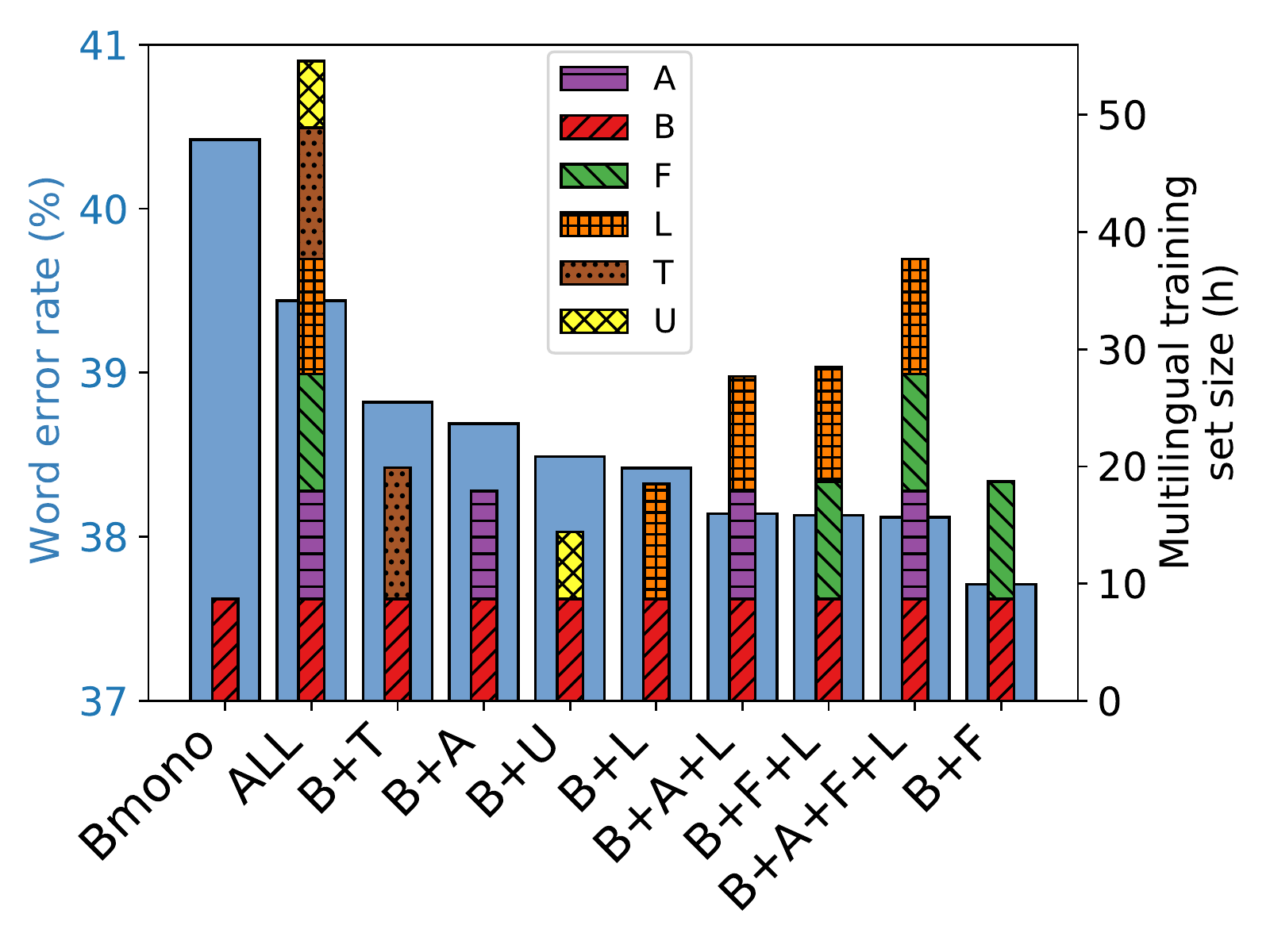} \\
\raggedleft(b) & \includegraphics[width=0.6\linewidth]{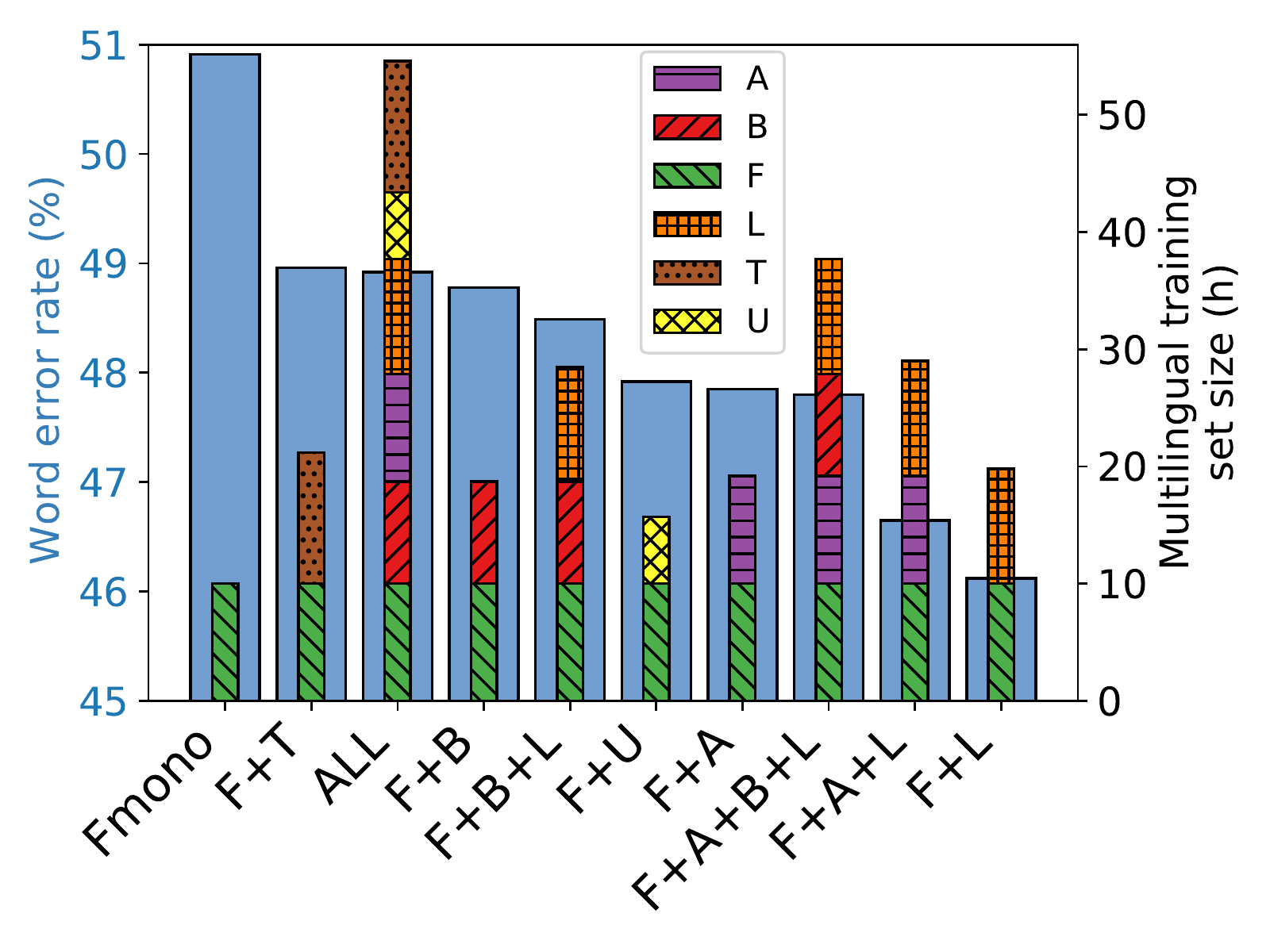} \\
\end{tabular}
\caption{
Word error rate results for the (a) Bambara and (b) Fulfulde ASR for the different multilingual training set combinations.
(Bmono: Monolingual Bambara; Fmono: Monolingual Fulfulde; A: Acholi; B: Bambara; F: Fulfulde; L: Luganda; T: Tunisian Modern Standard Arabic; U: Ugandan English.)
}
\label{fig:bam_ful_wer_results}
\end{figure}

The plots in Figure~\ref{fig:bam_ful_wer_results} depict the test set word error rates (WER) of the monolingual and multilingual systems for Bambara and Maasina Fulfulde.
The datasets used in these experiments have been described in Section~\ref{sec:corpora}.
In these plots, the wide blue bars indicate the test set word error rate (left vertical axis).
The narrow stacked bars indicate the sizes of the constituent corpora making up the training pool (right vertical axis).
Each colour represents a language, as indicated in the legend.

Firstly, we observe that the inclusion of any additional dataset, however small it is, leads to an improvement over the baseline monolingual system.
This indicates that, in our under-resourced setting, multilingual training is always beneficial.
Secondly, we observed that although the ALL set is able to outperform the monolingual baseline, it is by no means the best performing combination.
Hence a decision to simply pool all available out-of-language data is not the best strategy.
For Bambara, the best performing ASR system was trained in combination with Fulfulde and for the Fulfulde system the best result was obtained in combination with Luganda.
An interesting observation is that the inclusion of the Ugandan English dataset, which is the smallest among the six corpora (5.7 hours), results in a notable improvement in ASR performance (Bmono vs B+U and Fmono vs F+U).
In comparison, the inclusion of the Tunisian Arabic dataset, which is the largest among the six corpora (11.2 hours), resulted in a smaller improvement in ASR performance (WER for B+T$>$B+U and F+T$>$F+U).
This further affirms that the size of additional datasets cannot be considered a reliable indicator of the expected performance improvement.

The reasons for why one language combination results in better ASR performance than another are still not clear.
We have performed an analysis of the phonetic overlap between the languages, but this did not provide conclusive insight.
In particular, the phonetic overlap between the four Niger-Congo languages is fairly even, with Bambara sharing approximately 80\% of its phonetic inventory with the other three, while for Fulfulde it is approximately 68\%.
We suspect that there may be other factors at play, such as language characteristics besides common phonology, or similarities in recording and channel conditions.
This warrants further investigation.

\section{Conclusions}
\label{sec:conclusions}

We present first speech recognition systems for the two severely under-resourced Malian languages Bambara and Maasina Fulfulde.
These systems will be used by the United Nations to inform and support its humanitarian programmes in that country.
Because the datasets in the target languages are very small, extensive use has been made of multilingual acoustic modelling.
Since a number of datasets in other under-resourced languages were available for experimentation, some from related projects, a key question has been what the composition of the pool of data should be for best multilingual training.
We consider various combinations of up to six languages, whose corpora vary in size from  5.7 to 11.2 hours of speech,  and compare the performance of the resulting acoustic models in speech recognition experiments.
We found that, while including all six languages leads to improved speech recognition for both Bambara and Maasina Fulfulde, best performance is achieved when placing only one additional language in the pool.
For the case for Bambara, this additional language was Maasina Fulfulde, and its introduction leads to a relative word error rate reduction of 6.7\%, as opposed to a 2.4\% relative reduction achieved when pooling all six additional languages.
For the case of Maasina Fulfulde, best performance was achieved when adding only Luganda, leading to a relative improvement in word error rate of 9.4\% as opposed to a 3.9\% relative improvement observed when pooling all six languages.
We conclude that maximising the pool of multilingual data does not necessarily lead to the best-performing acoustic model for an under-resourced language, and that substantially better performance may be possible by a more judicious configuration of the pool of languages used for multilingual neural network training.

\section{Acknowledgements}

We would like to thank United Nations Global Pulse for collaboration and supporting this research.
We also gratefully acknowledge the support of NVIDIA corporation with the donation GPU equipment used during the course of this research, as well as the support of Council for Scientific and Industrial Research (CSIR), Department of Science and Technology, South Africa for provisioning us the Lengau CHPC cluster for seamlessly conducting our experiments.
We also gratefully acknowledge the support of Telkom South Africa.

\bibliographystyle{splncs04}

\bibliography{mybib}

\begin{thebibliography}{10}
\providecommand{\url}[1]{\texttt{#1}}
\providecommand{\urlprefix}{URL }
\providecommand{\doi}[1]{https://doi.org/#1}

\bibitem{Arnott-1970}
Arnott, D.W.: {The Nominal and Verbal Systems of Fula}. Clarendon Press, Oxford
  (1970)

\bibitem{arnott1974some}
Arnott, D.W.: {Some aspects of the study of Fula dialects}. Bulletin of the
  School of Oriental and African Studies, University of London  \textbf{37}(1),
   8--18 (1974)

\bibitem{barry2018proposal}
Barry, A., Barry, I., Constable, P., Glass, A.: Proposal to encode {ADLAM}
  nasalization mark for {ADLaM} script  (2018)

\bibitem{Boersma2021}
Boersma, P., Weenink, D.: {Praat: doing phonetics by computer [Computer
  program]. Version 6.1.39. Accessed: 2021-03-26}, \url{http://www.praat.org/}

\bibitem{Donaldson2017b}
Donaldson, C.: {Clear Language: Script, Register and the N'ko Movement of
  Manding-speaking West Africa}. Ph.D. thesis, University of Pennsylvania
  (2017)

\bibitem{Donaldson2017a}
Donaldson, C.: {Orthography, Standardization, and Register: The Case of
  Manding}. In: Lane, P., Costa, J., Korne, H.D. (eds.) {Standardizing Minority
  Languages: Competing Ideologies of Authority and Authenticity in the Global
  Periphery}, pp. 175--199. Routledge, 1st edn. (2017).
  \doi{https://doi.org/10.4324/9781315647722}

\bibitem{fagerberg1984}
Fagerberg-Diallo, S.: A Practical Guide and Reference Grammar to the Fulfulde
  of Maasina 1 \& 2. Joint Christian Ministry in West Africa, Jos, Nigeria
  (1984)

\bibitem{grezl2014adaptation}
Grézl, F., Karafiát, M., Veselý, K.: Adaptation of multilingual stacked
  bottle-neck neural network structure for new language. In:
  \textit{Proceedings of 2014 IEEE International Conference on Acoustics,
  Speech and Signal Processing} (ICASSP). Florence, Italy (2014)

\bibitem{heigold2013dnn}
Heigold, G., Vanhoucke, V., Senior, A., Nguyen, P., Ranzato, M., Devin, M.,
  Dean, J.: Multilingual acoustic models using distributed deep neural
  networks. In: \textit{Proceedings of 2013 IEEE International Conference on
  Acoustics, Speech and Signal Processing} (ICASSP). Vancouver, Canada (2013)

\bibitem{huang2013dnn}
Huang, J.T., Li, J., Yu, D., Deng, L., Gong, Y.: Cross-language knowledge
  transfer using multilingual deep neural network with shared hidden layers.
  In: \textit{Proceedings of 2013 IEEE International Conference on Acoustics,
  Speech and Signal Processing} (ICASSP). Vancouver, Canada (2013)

\bibitem{katzner2002languages}
Katzner, K., Miller, K.: The languages of the world. Routledge (2002)

\bibitem{Ko2015}
Ko, T., Peddinti, V., Povey, D., Khudanpur, S.: {Audio augmentation for speech
  recognition}. In: \textit{Proceedings of Interspeech} 2015. Dresden, Germany
  (2015)

\bibitem{menon2017radio}
Menon, R., Saeb, A., Cameron, H., Kibira, W., Quinn, J., Niesler, T.:
  {Radio-browsing for developmental monitoring in Uganda}. In:
  \textit{Proceedings of 2017 IEEE International Conference on Acoustics,
  Speech and Signal Processing} (ICASSP). New Orleans, USA (2017)

\bibitem{Osborn1993}
Osborn, D.W., Dwyer, D.J., Donohoe, J.I.J.: A Fulfulde (Maasina) - English -
  French Lexicon: A Root-Based Compilation Drawn from Extant Sources. Michigan
  State University Press (1993)

\bibitem{Padhi2020}
Padhi, T., Biswas, A., de~Wet, F., van~der Westhuizen, E., Niesler, T.:
  {Multilingual bottleneck features for improving ASR performance of
  code-switched speech in under-resourced languages}. In: \textit{Proceedings
  of the First Workshop on Speech Technologies for Code-switching in
  Multilingual Communities} (WSTCSMC). Shanghai, China (2020)

\bibitem{Park2019}
Park, D.S., Chan, W., Zhang, Y., Chiu, C.C., Zoph, B., Cubuk, E.D., Le, Q.V.:
  {SpecAugment: A Simple Data Augmentation Method for Automatic Speech
  Recognition}. In: \textit{Proceedings of Interspeech} 2019. Graz, Austria
  (2019)

\bibitem{povey2011kaldi}
Povey, D., Ghoshal, A., Boulianne, G., Burget, L., Glembek, O., Goel, N.,
  Hannemann, M., Motlicek, P., Qian, Y., Schwarz, P., et~al.: The {K}aldi
  speech recognition toolkit. In: \textit{Proceedings of 2011 IEEE Workshop on
  Automatic Speech Recognition and Understanding} (ASRU). Hawaii, USA (2011)

\bibitem{Povey2016}
Povey, D., Peddinti, V., Galvez, D., Ghahremani, P., Manohar, V., Na, X., Wang,
  Y., Khudanpur, S.: {Purely sequence-trained neural networks for ASR based on
  lattice-free MMI}. In: Proceedings of Interspeech (2016)

\bibitem{Saeb2017}
Saeb, A., Menon, R., Cameron, H., Kibira, W., Quinn, J., Niesler, T.: {Very low
  resource radio browsing for agile developmental and humanitarian monitoring}.
  In: \textit{Proceedings of Interspeech} 2017. Stockholm, Sweden (2017)

\bibitem{schultz2006multilingual}
Schultz, T., Kirchhoff, K.: Multilingual speech processing. Elsevier (2006)

\bibitem{schultz2001language}
Schultz, T., Waibel, A.: Language-independent and language-adaptive acoustic
  modeling for speech recognition. Speech Communication  \textbf{35}(1-2),
  31--51 (2001)

\bibitem{Stolcke2002}
Stolcke, A.: Srilm - an extensible language modeling toolkit. In:
  \textit{Proceedings of Interspeech} 2002. Denver, Colorado (2002)

\bibitem{Tachbelie2020}
Tachbelie, M.Y., Abate, S.T., Schultz, T.: {Development of multilingual ASR
  using globalphone for less-resourced languages: The case of Ethiopian
  languages}. In: \textit{Proceedings of Interspeech} 2020. Shanghai, China
  (2020)

\bibitem{vesely2012language}
Vesel{\`y}, K., Karafi{\'a}t, M., Gr{\'e}zl, F., Janda, M., Egorova, E.: The
  language-independent bottleneck features. In: \textit{Proceedings of 2012
  IEEE Spoken Language Technology Workshop} (SLT). Miami, USA (2012)

\bibitem{Vydrine2015}
Vydrine, V.: {Manding-English Dictionary: Maninka, Bamana Vol. 1}. MeaBooks
  Inc. (2015)

\end{thebibliography}

\end{document}